# Cooperative Jahn-Teller dynamics of boron clusters in the infrared conductivity of heavy fermion metal $CeB_6$ and unconventional superconductor $ZrB_{12}$

Gennady Komandin[a], Elena Zhukova[b], Boris Gorshunov[b], Andrey Azarevich[a,b], Andrey Muratov[c], Yurii Aleshchenko[c], Nikolay Sluchanko[a,]*

[a]Prokhorov General Physics Institute of Russian Academy of Sciences, 119991 Moscow, Vavilov str.,38, Russia
[b]Laboratory of Terrahertz spectroscopy, Center for Photonics and 2D Materials, Moscow Institute of Physics and Technology (national Research University), Dolgoprudny, Moscow Region, Russia
[c]Lebedev Physical Institute of RAS, 119991 Moscow, Leninskiy Av. 53, Russia

*e-mail: nes@lt.gpi.ru

**Abstract**
A thorough study of the wide-range (40-35000 cm$^{-1}$) dynamic conductivity and permittivity spectra of the archetypal heavy fermion metal $CeB_6$ and unconventional superconductor $ZrB_{12}$ was carried out at room temperature. Both the Drude-type components and overdamped excitations were separated and analyzed. An additional absorption band observed above 200 cm$^{-1}$ was attributed to the cooperative Jahn-Teller dynamics of the boron complexes in $CeB_6$ and $ZrB_{12}$. It was shown that nonequilibrium electrons participating in the formation of the collective JT modes dominate in charge transport, and fraction of Drude-type electrons does not exceed 37% in $CeB_6$ and 23% in $ZrB_{12}$. We discuss also the additional Drude-type component in $ZrB_{12}$ in terms of far-infrared conductivity from the sliding charge density wave, and suggest the localized mode scenario reconciles the strong difference between the number of conduction electrons obtained from the Hall effect and optical sum rule analysis in $CeB_6$.

**Introduction**
Even though studies of the properties and complex magnetic phase diagram of the archetypal heavy-fermion (HF) metal $CeB_6$ began over 55 years ago, this unusual antiferromagnet with a simple cubic boron-rich lattice (see inset in Fig. 1a) has remained the subject of ongoing experimental and theoretical research to this day. In particular, the nature of low-temperature magnetic phase II (AFQ) remains a topic of debate (see, for example, references [2–10]). The models proposed to describe the homogeneous single-phase state (see the review in [11] and the references cited therein) are inadequate for $CeB_6$, which exhibits electron phase separation. Indeed, it was argued in [12] that, rather than exhibiting single-ion behavior, $CeB_6$ exhibits the Griffiths phase scenario involving nanoscale magnetic clusters of Ce ions over an extra-wide temperature range of 5–800 K. This corresponds to an absolute record value of the Griffiths temperature (TG) of >800 K for disordered systems. In addition, a power-law dependence of resistivity $\rho(T) \sim T^{-0.4}$ was deduced in this metal within the temperature range of 8–90 K [12]. This points to a scenario that is opposite to the Kondo-lattice one and favors the weak-localization regime of charge carriers located inside the nanoscale Ce clusters in this strongly correlated electron system. Moreover, precise X-ray diffraction (XRD) studies of high-quality single crystals revealed electron and lattice instabilities in $CeB_6$ in the temperature range of 30–500 K. These instabilities are responsible for the emergence of dynamic charge stripes and sub-structural charge density waves (s-CDWs) [11–13], as well as the formation of ferromagnetic nanoscale droplets (ferrons) in the mysterious AFQ magnetic phase [11].

Similar nanoscale heterogeneity has also been established in the magnetic hexaboride $GdB_6$ [14] and the non-magnetic reference compound $LaB_6$ [12, 15]. The latter compound exhibits record thermionic emission, which is currently utilized in numerous practical applications [16]. Studies of the wide-band dynamic conductivity of $LaB_6$ single crystals with various $^{10}B$ and $^{11}B$ isotope contents [17], as well as substitutional solid solutions of $Gd_xLa_{1-x}B_6$ [18], suggest that only a small fraction of conduction electrons behave as Drude-type charge carriers. Around 70% of electrons, meanwhile, are involved in collective oscillations of electron density coupled to vibrations of the unstable Jahn–Teller rigid covalent boron cage and the rattling modes of the loosely bound heavy La (Gd) ions. It has been suggested that it is the non-equilibrium (hot) conduction electrons that determine the extraordinarily low work function of thermionic emission in $LaB_6$ [17]. Additionally, the collective Jahn–Teller (JT) mode (see [19] for more details) triggers periodic changes in the hybridization of the $5d$-$2p$ conduction band states, leading to nanoscale electron phase separation (dynamic charge stripes). Detailed dynamic conductivity studies of $CeB_6$ have not yet been conducted, so characterizing conduction electrons in the archetypal HF hexaboride is the focus of the present research.

$ZrB_{12}$ is another high boride presented in the study and is believed to be a model superconductor for various classes of high-$T_c$ (HTSC) compounds, including HTSC cuprates with collective dynamics of $O_6$ octahedrons centered on Cu-ions and polyhydrides with labile $H_n$ clusters. These are currently the highest-temperature superconductors, with transition temperatures of ~250–270 K (see, for example, references [21–23]). The two-gap superconductivity was established in $ZrB_{12}$ in various experiments (see [20] for a recent review). In addition, the formation of grids of dynamic charge stripes and sub-structural charge density waves (s-CDWs) was detected in [24, 20], alongside the pseudo-gap state observed in $ZrB_{12}$ during photoemission spectroscopy experiments [25]. These findings are usually considered as the fingerprints of the unconventional HTSC, suggesting common features of unconventional superconductivity in $ZrB_{12}$, and in cuprates and Fe-based pnictides. A new *vibron-plasmon-phonon superconducting pairing mechanism* was proposed in [20], which may be common for various classes of HTSCs, including cage-cluster polyhydrides $(La,Y)H_n$ (n=10) with *fcc* crystal structure (space group $Fm\bar{3}m$) and the highest $T_c \geq 250$ K discovered at pressures up to 170 GPa.

When discussing the nature of the *fcc* lattice and electronic structure instabilities in rare earth (RE) and transition metal (TM) dodecaborides ($RB_{12}$), it is worth noting that the cooperative Jahn–Teller (JT) lability of $B_{12}$ clusters (ferrodistortive effect) should be considered as the source of the very small static deformation of *fcc* lattice (see inset in Fig. 1b), which was observed in $LuB_{12}$ [19, 26] and $ZrB_{12}$ [20], in conjunction with the strong dynamic JT effect that develops within the rigid boron cage. It has been suggested that the dynamic component of the JT effect in $RB_{12}$ is responsible for the periodic changes in the $4d(5d)$-$2p$ hybridization of the TM (RE) and boron electron states, which lead to modulation of the conduction band width and the formation of dynamic charge stripes. These stripes can be considered as charge-rich channels with fluctuating electron density. Dynamic conductivity measurements have been carried out on $LuB_{12}$ with various $^{10}B$ and $^{11}B$ isotope content [27], concluding that 70–80% of conduction electrons are involved in forming collective JT excitations (hot charge carriers). Thus, the present study aims to separate and analyze in detail the various contributions to the dynamic conductivity of unconventional superconductor $ZrB_{12}$ and also heavy fermion metal $CeB_6$.

**Experimental details**

High quality $CeB_6$ and $ZrB_{12}$ single crystals were grown by inductive zone melting technique, as described in [28], and carefully characterized by the XRD, chemical microanalysis and charge transport measurements. For optical measurements, round samples ~5-6 mm in diameter were used with plane (within ±1 μm) polished surface that was etched in the boiling $HNO_3+H_2O$ solution to remove the surface layer with possible structural distortions [28]. Infrared (IR) reflectivity spectra $R(v)$ were measured using Bruker Vertex 80V Fourier-transform spectrometer at frequencies $v$=40-8000 cm$^{-1}$. Gold film deposited on the glass substrate was used as a reference mirror. The real and imaginary parts of dielectric permittivity and dynamic conductivity at frequencies 3700 – 35000 cm$^{-1}$ were directly measured using the J.A. Woollam V-VASE ellipsometer. Measurements with the radiation spot diameter of 2 mm were provided with angles of incidence 65, 70 and 75 degrees. From the ellipsometry data, reflection coefficient spectra were calculated using standard Fresnel expressions and merged with the reflectivity spectra measured in the IR band. The data from [29] (CeB$_6$) and [30] (ZrB$_{12}$) were used to extend the reflectivity spectra up to ~400000 cm$^{-1}$. The so-obtained broad-band spectra of reflection coefficient were processed as described below. Direct current (DC) conductivity and Hall resistivity of the same crystals were measured using five-terminal Van-der-Pauw scheme. All measurements were performed at room temperature.

**Results and discussion**

Dots in Figures 1a and 1b present the measured broad-band reflectivity spectra of CeB$_6$ and ZrB$_{12}$ crystals. The spectra look typically metallic (see, e.g., [31]), displaying pronounced plasma edges (plasma minima) at a frequency of $\frac{\nu_{pl}}{\sqrt{\varepsilon_\infty}} \approx 17000$ cm$^{-1}$ for both CeB$_6$ and ZrB$_{12}$ (here $\varepsilon_\infty$ is the high-frequency dielectric permittivity and $\nu_{pl}$ is the plasma frequency). Below $\approx 1000$ cm$^{-1}$, the reflectivity increases towards low frequencies approaching 100%. Peaks seen above 20000 cm$^{-1}$ are caused by electronic interband transitions [29, 30].

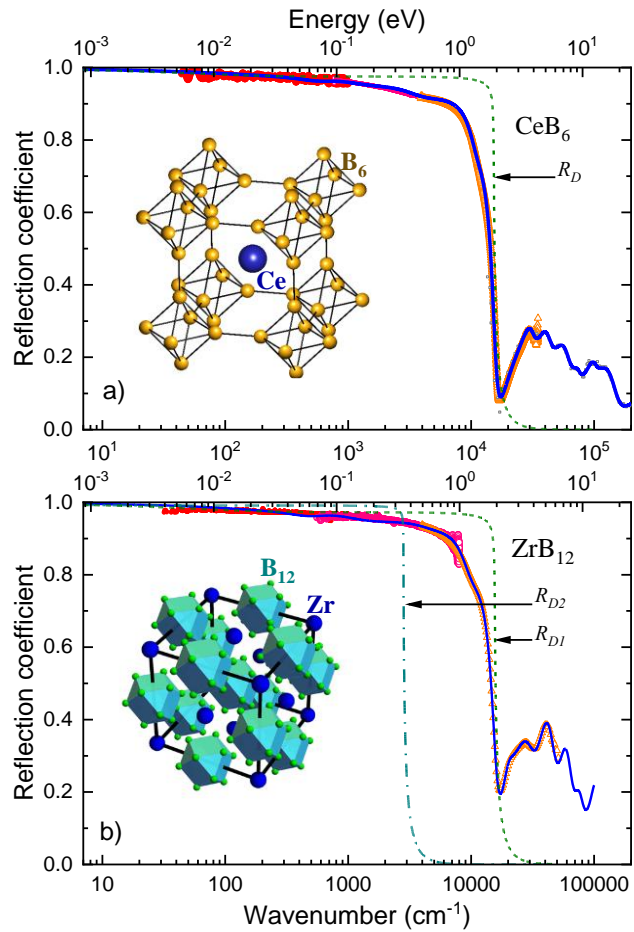

**Figure 1.** (Color online). Room temperature reflection coefficient spectra of CeB$_6$ (a) and ZrB$_{12}$ (b) single crystals. Dots show experimental data obtained using Fourier-transform spectrometer and ellipsometer. Open symbols above 20000 cm$^{-1}$ correspond to high-frequency reflectivity data from [29] (a) and [30] (b). Blue solid lines show the results of fitting the spectra using the Drude term, Eq. (1), for the free charge carrier response and Lorentzians, Eq. (2), responsible for absorption resonances. Dashed lines (R$_{D1}$) show best fits of the spectra that can be obtained using the Drude conductivity term alone with $\sigma_{DC}$=39000 $\Omega^{-1}$ cm$^{-1}$, $\gamma_D$=400 cm$^{-1}$ for CeB$_6$ and $\sigma_{DC1}$=35000 $\Omega^{-1}$ cm$^{-1}$, $\gamma_{D1}$=381 cm$^{-1}$ for ZrB$_{12}$. Dash-dotted line in (b) (R$_{D2}$) shows the contribution corresponding to an extra Drude conductivity term (s-CDW, see text) with $\sigma_{DC2}$=17640 $\Omega^{-1}$ cm$^{-1}$, $\gamma_{D2}$=62 cm$^{-1}$ for ZrB$_{12}$. Insets in panels (a) and (b) present the crystal structure of CeB$_6$ and ZrB$_{12}$, respectively.

Following the procedure used earlier to analyze the spectra of RE and TM high-borides LuB$_{12}$ [27], Tm$_{1-x}$Yb$_x$B$_{12}$ [32-34], Gd$_x$La$_{1-x}$B$_6$ [17-18] and YB$_6$-YbB$_6$ [35], we have processed the broadband reflectivity spectra together with the directly measured spectra of the real and imaginary parts of permittivity. The Drude conductivity model was used to describe the response of free carriers [31,36]

$$\sigma_D(\nu) = \sigma_{DC}/(1 - i\nu/\gamma_D), \qquad (1)$$

where $\sigma_{DC}$ is the direct current conductivity and $\gamma_D$ is the charge-carrier scattering rate. The broad and narrow spectral features were modelled with a set of Lorentzian terms

$$\sigma_j^*(\nu) = 0.5 f_j \nu / \left(\nu\gamma_j + i(\nu_{0j}^2 - \nu^2)\right), \qquad (2)$$

where $\nu_{0j}$ is the resonance frequency, $f_j = \Delta\varepsilon_j \nu_{0j}^2$ is the oscillator strength, $\Delta\varepsilon_j$ is the dielectric contribution and $\gamma_j$ is the damping constant of the *j*-th Lorentzian. For ZrB$_{12}$, in addition to the Drude and four Lorentzian terms, we had to introduce an extra Drude term to describe the sliding CDW contribution [20]. The spectra below $\approx 20000$ cm$^{-1}$ were described using the minimal set of four Lorentzians (L$_1$–L$_4$). The higher-frequency inter-band transitions were also modelled using Lorentzians and will not be discussed here.

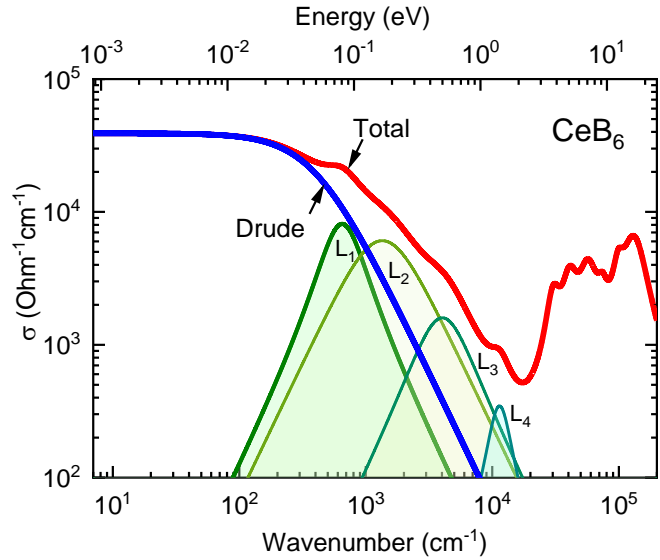

**Figure 2.** (Color online). Room temperature spectrum of real part of conductivity of CeB$_6$ crystal (red solid line), obtained by least square fitting of the reflection coefficient spectrum using Drude conductivity Eq. (1) and Lorentzian terms Eq. (2), as described in the text. The Drude and Lorentzian L$_1$-L$_4$ contributions are shown separately.

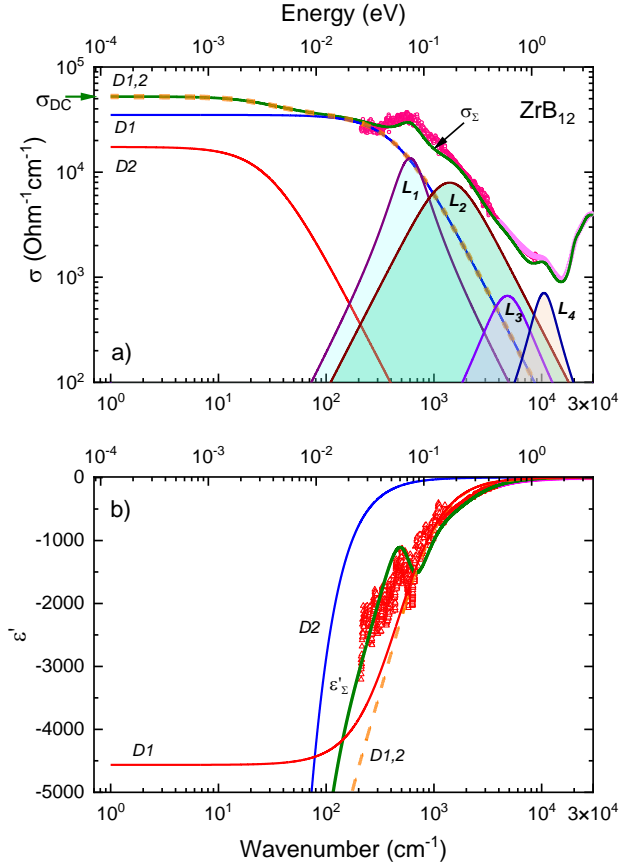

**Figure 3.** (Color online). Room temperature spectra of real parts of (a) conductivity and (b) dielectric permittivity of ZrB$_{12}$ (dots). $\sigma_\Sigma(\nu)$ and $\varepsilon'_\Sigma(\nu)$ are obtained by least square fitting of the reflection coefficient spectra using Drude conductivity components D$_1$, D$_2$ Eq. (1) and Lorentzian terms L$_1$-L$_4$ Eq. (2). Arrow denotes $\sigma_{DC}$ value measured by van der Pauw technique; D$_1$, D$_2$ and D$_{1,2}$ curves show the Drude terms which correspond to the parameters presented in the Table 2, and their sum.

The results obtained for CeB$_6$ and ZrB$_{12}$ are shown by the blue lines in Figure 1. The corresponding optical conductivity and permittivity spectra, showing the contributions of all terms separately, are presented in Figures 2 and 3. Tables 1 and 2 show the obtained parameters of the Drude and Lorentzian terms.

The observed absorption bands (L$_1$-L$_4$) have rather unusual characteristics. Firstly, these bands are clearly visible in both the reflectivity and conductivity spectra, as they are not fully obscured by charge carriers. Secondly, they have rather large values of the oscillator strengths $f_j \sim (0.1–1.1) \times 10^9$ cm$^{-2}$, the dielectric contributions $\Delta\varepsilon_{1,2} \sim 400$ –1200 and the damping constants $\gamma_j/\nu_{0j} \sim 0.5 - 1.5$ (see Tables 1 and 2). This implies that the relevant absorption mechanisms cannot be associated with inter-band transitions. Similar to the conclusions drawn in references [17–18, 27, 32–35], we assume that the origin of the observed overdamped excitations is related to the Jahn–Teller (JT) instability of the B$_6$ and B$_{12}$ complexes of natural boron (which is a mixture of isotopes $^{10}$B and $^{11}$B, approximately in a 1:4 ratio) in CeB$_6$ and ZrB$_{12}$, respectively, leading to the emergence of both static and cooperative dynamic JT effects [19]. We suggest that this kind of non-adiabatic mechanism, which launches both the collective overdamped modes and the electronic instability, provides the most natural interpretation of the anomalies observed in the $\sigma(\nu)$ spectra [19].

The Drude and Lorentzian parameters were used to estimate the concentration of the Drude-type conduction electrons and hot charge carriers involved in the cooperative JT dynamics. Specifically, taking for Drude electrons in ZrB$_{12}$ the value of effective mass $m*=0.8m_0$ ($m_0$ is the free electron mass) as found in the de Haas-van Alphen measurements [37], and using the relations for charge carriers plasma frequency $\nu_{pl} = [n_D e^2/\pi m^*]^{1/2}$ ($n_D$ is the concentration of Drude-type electrons and $e$ – their charge) and sum of oscillator strengths of Lorentzians $f = \sum f_j = \sum \Delta\varepsilon_j v_{0j}^2 = n_{peak} e^2 (\pi m)^{-1}$, we obtain concentrations $n_D$ of carriers participating in the Drude conductivity and $n_{peak}$ of electrons involved in the formation of the overdamped excitations. Since the electron-phonon scattering in the collective JT mode is very strong, the mean free path $l$ of hot electrons must be small enough and satisfy the condition $m* \approx m_0$. The same value $m* \sim m_0$ can be taken as a lower estimate of the effective mass of different types' electrons in the case of heavy fermion metal CeB$_6$, which is already in the Griffiths phase regime at room temperature [12]. The concentrations $n_D$ and $n_{peak}$ found for CeB$_6$ and ZrB$_{12}$ are presented in the Tables 1 and 2. It is worth noting that the values of the total concentration of electrons $n_{opt} = n_D + n_{peak}$ are very similar in CeB$_6$ ($n_{opt} \approx 2.7 \cdot 10^{22}$ cm$^{-3}$) and ZrB$_{12}$ ($n_{opt} \approx 2.73 \cdot 10^{22}$ cm$^{-3}$), even though CeB$_6$ is a monovalent and ZrB$_{12}$ is divalent. Strongly different are the relative (normalized to the number of Ce(Zr) ions) concentrations, $N_{eff}$(CeB$_6$)= $n_{opt}$ /$n_{Ce} \approx 1.93$ and $N_{eff}$(ZrB$_{12}$) $\approx 2.8$. A close value $N_{eff}$(ZrB$_{12}$) $\approx 2.9$ was estimated from integration of the optical conductivity spectra recorded at several temperatures [38]. The fraction of nonequilibrium charge carriers (hot electrons) involved in the formation of the collective JT excitations in CeB$_6$ is $n_{peak}/n_{opt} \approx 63\%$ and reaches 77% in ZrB$_{12}$ (see Tables 1 and 2). These values are close to those one detected in LuB$_{12}$ [27], Tm$_{1-x}$Yb$_x$B$_{12}$ [32-34], Gd$_x$La$_{1-x}$B$_6$ [17-18] and YB$_6$-YbB$_6$ [35]. Note that the charge density wave contribution to dynamic conductivity proposed here (D$_2$ in Figure 3 and Table 2) comprises only 2.7% of the content of charge carriers, providing a small correction to the IR conductivity spectrum.

**Table 1.** Parameters of the Drude (D) and Lorentzian (L$_1$-L$_4$) components obtained by processing the spectra of CeB$_6$ single crystal: $\sigma_{DC}$ is the static conductivity, $\gamma_D$ is the charge-carrier scattering rate, $v_{pl}$ is the plasma frequency, $n_D$ and $n_{peak}$ are the charge carrier concentration of Drude-type and hot electrons, respectively, $\Delta\varepsilon$ is the dielectric contribution, $\gamma$ is the damping constant and $f$ is the oscillator strength.

| Drude term | | | |
|---|---|---|---|
| $\sigma_{DC}$ Ohm$^{-1}$cm$^{-1}$ | $\gamma_D$ cm$^{-1}$ | $v_{pl}$ cm$^{-1}$ (Hz) | $n_D$ cm$^{-3}$ |
| 39000 | 400 | 30600 (9.18·10$^{14}$) | 1.0·10$^{22}$ |

| Lorentzians | $\Delta\varepsilon$ | $v_{pl}$ cm$^{-1}$ | $\gamma$ cm$^{-1}$ | $f$ cm$^{-2}$ (Hz$^{-2}$) | $n_{peak}$ cm$^{-3}$ |
|---|---|---|---|---|---|
| L$_1$ | 600 | 650 | 520 | 2.5·10$^8$ (2.28·10$^{29}$) | 2.8·10$^{21}$ |
| L$_2$ | 400 | 1350 | 2000 | 7.3·10$^8$ (6.5·10$^{29}$) | 8.1·10$^{21}$ |
| L$_3$ | 23 | 4030 | 4240 | 3.7·10$^8$ (3.36·10$^{29}$) | 4.5·10$^{21}$ |
| L$_4$ | 0.8 | 11400 | 5000 | 1·10$^8$ (0.93·10$^{29}$) | 1.2·10$^{21}$ |
| | | | | | $n_{peak} \approx 1.7 \cdot 10^{22}$ |

**Table 2** Parameters of the Drude (D$_1$-D$_2$) and Lorentzian (L$_1$-L$_4$) components obtained by processing the spectra of ZrB$_{12}$ single crystal: $\sigma_{DC}$ is the static conductivity, $\gamma_D$ is the charge-carrier scattering rates, $v_{pl}$ - the plasma frequency, $n_D$ and $n_{peak}$ are the charge carrier concentration of Drude-type and hot electrons, respectively, $\Delta\varepsilon$ is the dielectric contribution, $\gamma$ is the damping constant and $f$ is the oscillator strength.

| Drude terms | $\sigma_{DC}$ Ohm$^{-1}$cm$^{-1}$ | $\gamma_D$ cm$^{-1}$ | $v_{pl}$ cm$^{-1}$ (Hz) | $n_D$, cm$^{-3}$ |
|---|---|---|---|---|
| D1 | 35000 | 381 | 28396 (8.5·10$^{14}$) | 6.3·10$^{21}$ |
| D2 | 17640 | 62 | 8142 (2.4·10$^{14}$) | 0.74·10$^{21}$ |

Lorentzians

|    | $\Delta\varepsilon$ | $\nu$, cm$^{-1}$ | $\gamma$, cm$^{-1}$ | $f$ cm$^{-2}$ (Hz$^{-2}$) | $n_{\text{peak}}$ cm$^{-3}$ |
|----|------|------|------|------|------|
| L$_1$ | 1160 | 600   | 450  | $4.2\cdot10^8$ ($3.8\cdot10^{29}$) | $4.7\cdot10^{21}$ |
| L$_2$ | 560  | 1400  | 2000 | $1.1\cdot10^9$ ($9.9\cdot10^{29}$) | $12.2\cdot10^{21}$ |
| L$_3$ | 9    | 4800  | 4500 | $2.1\cdot10^8$ ($1.9\cdot10^{29}$) | $2.31\cdot10^{21}$ |
| L$_4$ | 2.4  | 10500 | 5700 | $2.6\cdot10^8$ ($2.4\cdot10^{29}$) | $2.95\cdot10^{21}$ |
|    |      |       |      |      | $n_{\text{peak}} \approx 2.1\cdot10^{22}$ |

The two components $\sigma_D$ and $\sigma_{IR}$ present in the optical conductivity spectra of ZrB$_{12}$ have been found earlier in the light and heavy RE hexaborides RB$_6$ [39], and the IR absorption peak was detected in CeB$_6$ above 2400 cm$^{-1}$ (~0.3 eV). The authors of Ref. [39] observed an absorption structure in the IR conductivity spectra around 0.6 eV for all RB$_6$. In the case of CeB$_6$, the $\sigma(\nu)$ spectrum was analyzed in the range below 0.3 eV, considering only the Drude-type conductivity component. As a result, the spectral weights of the Lorentzians L$_1$ and L$_2$ detected in the present study (see Figure 2) were attributed in [39] to the Drude contribution, leading to an incorrect separation of components and wrong evaluation of concentrations $n_D$ and $n_{IR}$ (=$n_{\text{peak}}$).

The total concentration $n_{\text{opt}} = n_D + n_{\text{peak}} \approx 2.73\cdot10^{22}$ cm$^{-3}$ found from the above analysis of the optical spectra of ZrB$_{12}$ (Figure 3 and Table 2) is comparable to the concentration $n_H = 1/(R_H e) \approx 2.4$-$2.5\cdot10^{22}$ cm$^{-3}$ obtained in [20] from the Hall effect measurements performed on the same crystals ($R_H$ is the Hall coefficient). The situation is dramatically different in the case of the HF metal CeB$_6$, for which $n_H \approx 1.4\cdot10^{22}$ cm$^{-3}$ was obtained. This value is approximately equal to the number of Ce-ions $n_{\text{Ce}}$ in the monovalent metal but about half the total value $n_{\text{opt}}$ estimated above. In order to reconcile these $n_H$ and $n_{\text{opt}}$ values, it is necessary to propose an effective mass of $m^* \sim 0.5 m_0$, which is inconsistent with the generally accepted heavy-fermion scenario for CeB$_6$ [11]. This contradiction was first highlighted by Kasuya et al. in [39], where a formation of $4f$-$5d$ exciton was discussed for RB$_6$. However, the exciton mechanism seems questionable, since only very small deviation from the Ce$^{3+}$ valence ($\upsilon \leq 3.05$, see [40], for example) was deduced in this HF metal with fast $4f$-$5d$ charge fluctuations. In light of the weak localization observed in charge transport measurements at temperatures of 8–90 K [11, 12], it is reasonable to suggest that some of the overdamped excitations identified in the room-temperature conductivity spectrum of CeB$_6$ should be regarded as localized modes that do not participate in charge transport. Such kind of absorption resonances were discovered recently in the optical conductivity spectra of the strongly correlated narrow-gap semiconductors Tm$_{1-x}$Yb$_x$B$_{12}$ with metal-insulator transition (MIT) [33]. The broad absorption band in the conductivity spectrum centered at about 1800 cm$^{-1}$ was observed for Tm$_{0.19}$Yb$_{0.81}$B$_{12}$, and it changes only slightly during the MIT when cooling the crystal from 300 K to 10 K [33]. In [33], the authors associate its origin with the vibrationally coupled localized states near the bottom of the conduction band [34]. The position, width and dielectric contribution of these absorption peaks in CeB$_6$ (see Figure 2) and Tm$_{1-x}$Yb$_x$B$_{12}$ [33] are very similar, arguing in favor of the proposed interpretation. Our low-temperature optical measurements are in progress, which will shed more light on the nature of weak localization phenomenon and on the mechanism underlying formation of these localized states.

**Conclusions.**

The systematic room temperature study of the wide-range (40-35000 cm$^{-1}$) spectra of dynamic conductivity and permittivity of the archetypal heavy fermion compound CeB$_6$ and unconventional superconductor ZrB$_{12}$ was carried out. The detailed analysis of the dynamic conductivity developed here allowed to separate Drude-type components and to discover overdamped excitations connected with cooperative Jahn-Teller instability of the B$_6$ and B$_{12}$ complexes in CeB$_6$ and ZrB$_{12}$, respectively. We estimate the concentration of charge carriers associated with these conductivity channels and show that fraction of nonequilibrium (hot) electrons involved in the formation of the collective JT excitations is about 63% in CeB$_6$ and 77% in ZrB$_{12}$. We propose interpretation of the origin of the additional Drude-type component discovered in ZrB$_{12}$ in terms of contribution to the far-IR conductivity from sliding charge density wave. We suggest the localized modes scenario to reconcile a twofold difference in the charge carriers' concentrations obtained from the Hall effect and from the optical sum rule analysis in the heavy fermion metal $CeB_6$.

**Acknowledgements**
The work was partly performed using the equipment of the Shared Facilities Center of Lebedev Physical Institute of RAS. Infrared experiments and the analysis of spectra were supported by the Russian Science Foundation, projects No. 25-42-00058 and No. 25-79-30006. The authors are grateful to S.V. Demishev, V.V. Glushkov, N. B. Bolotina, S. Gabani and K. Flachbart for useful discussions.

**References**
1. Yu.B. Paderno, S. Pokrzywnicki, and B. Stalinski, Magnetic properties of some rare earth hexaborides, Phys. status solidi (b) **24**, K73 (1967).
2. A. S. Cameron, G. Friemel, and D. S. Inosov, Multipolar phases and magnetically hidden order: review of the heavy−fermion compound $Ce_{1-x}La_xB_6$, Rep. Prog. Phys. **79**, 66502 (2016).
3. S.V. Demishev, V.N. Krasnorussky, A.V. Bogach, V.V. Voronov, N.Yu. Shitsevalova, V.B. Filipov, V.V. Glushkov, and N.E. Sluchanko, Electron nematic effect induced by magnetic field in antiferroquadrupole phase of $CeB_6$, Sci. Rep. **7**, 17430 (2017).
4. H. Jang, G. Friemel, J. Ollivier, A.V. Dukhnenko, N.Y. Shitsevalova, V.B. Filipov, B. Keimer, and D.S. Inosov, Intense low−energy ferromagnetic fluctuations in the antiferromagnetic heavy-fermion metal $CeB_6$, Nat. Mater. **13**, 682 (2014).
5. S.V. Demishev, A.V. Semeno, A.V. Bogach, N.A. Samarin, T.V. Ishchenko, V.B. Filipov, N.Yu. Shitsevalova, and N.E. Sluchanko, Magnetic spin resonanse in $CeB_6$, Phys. Rev. B **80** 245106 (2009).
6. A.V. Semeno, M.I. Gilmanov, A.V. Bogach, V.N. Krasnorussky, A.N. Samarin, N.A. Samarin, N.E. Sluchanko, N.Yu. Shitsevalova, V.B. Filipov, V.V. Glushkov, and S.V. Demishev, Magnetic resonance anisotropy in $CeB_6$: An entangled state of the art, Sci. Rep **6**, 39196 (2016).
7. A. Schenck, F.N. Gygax, and S. Kunii, Field−induced magnetization distribution and antiferroquadrupolar order in $CeB_6$, Phys. Rev. Lett., **89**, 037201 (2002).
8. A. Schenck, F.N. Gygax, G. Solt, O. Zaharko, and S. Kunii, Temperature and field dependence of the order parameter in the antiferroquadrupolar phase of $CeB_6$ from $\mu+$ Knight shift, Phys. Rev. Lett., **93**, 257601 (2004).
9. N.E. Sluchanko, A.V Bogach, V.V Glushkov, S.V Demishev, V.Yu. Ivanov, M.I. Ignatov, A.V Kuznetsov, N.A. Samarin, A.V Semeno, and N.Yu. Shitsevalova, Enhancement of band magnetism and features of the magnetically ordered state in the $CeB_6$ compound with strong electron correlations, JETP **104,** 120 (2007).
10. G. Friemel, Y. Li, A.V Dukhnenko, N.Yu. Shitsevalova, N.E. Sluchanko, A. Ivanov, V. B. Filipov, B. Keimer, and D. S. Inosov, Resonant magnetic exciton mode in the heavy-fermion antiferromagnet $CeB_6$, Nat. Commun. **3**, 830 (2012).
11. A.N. Azarevich, O.N. Khrykina, N.B. Bolotina, V.G. Gridchina, A.V. Bogach, S.V. Demishev, V.N. Krasnorussky, S. Yu. Gavrilkin, A.Yu. Tsvetkov, N.Yu. Shitsevalova, V.V. Voronov, K.I. Kugel, A.L. Rakhmanov, S. Gabáni, K. Flachbart, N.E. Sluchanko, Evidence for spin droplets (ferrons) formation in the heavy fermion metal $CeB_6$ with dynamic charge stripes, Solid State Sci. **167**, 107990 (2025).
12. O.N. Khrykina, N.B. Bolotina, V.G. Gridchina, A.N. Azarevich, A.V. Bogach, K.M. Krasikov, N.Yu. Shitsevalova, V.B. Filipov, and N.E. Sluchanko, Evidence for nanosized magnetic clusters of Ce ions in the archetypal heavy fermion metal $CeB_6$, J. Alloys Compd. **970**, 172527 (2024).
13. O.N. Khrykina, N.B. Bolotina, V.M. Gridchina, A.N. Azarevich, K.M. Krasikov, N.Yu. Shitsevalova, V.B. Filipov, S.Yu. Gavrilkin, A.Yu. Tsvetkov, and N.E. Sluchanko, Electronic phase transitions in heavy-fermion $CeB_6$ compound, JETP Lett. **119**, 144 (2024).

14. A.P. Dudka, O.N. Khrykina, N.B. Bolotina, N.Yu. Shitsevalova, V.B. Filipov, M.A. Anisimov, S. Gábani, K. Flachbart, and N.E. Sluchanko, Quantum diffusion regime of charge transport in $GdB_6$ caused by electron and lattice instability, Phys. Rev. B **100**, 205103 (2019).
15. A.N. Azarevich, A.V. Bogach, O.N. Khrykina, N.B. Bolotina, V.M. Gridchina, N.Yu. Shitsevalova, S.Yu. Gavrilkin, A.Yu. Tsvetkov, S. Gabáni, K. Flachbart, A.V. Kuznetsov, and N.E. Sluchanko, Localized superconductivity in $LaB_6$ hexaboride with dynamic charge stripes, JETP Lett. **119**, 934 (2024).
16. M. Trenary, Surface science studies of metal hexaborides, Sci. Technol. Adv. Mater. **13**, 023002 (2012).
17. E.S. Zhukova, B.P. Gorshunov, M. Dressel, G.A. Komandin, M.A. Belyanchikov, Z.V. Bedran, A.V. Muratov, Y.A. Aleshchenko, M.A. Anisimov, N.Yu. Shitsevalova, A.V. Dukhnenko, V.B. Filipov, V.V. Voronov, and N.E. Sluchanko, Boron $^{10}B$–$^{11}B$ isotope substitution as a probe of the mechanism responsible for the record thermionic emission in $LaB_6$ with the Jahn–Teller instability, JETP Lett. **110**, 79 (2019).
18. E.S. Zhukova, B.P. Gorshunov, G.A. Komandin, L.N. Alyabyeva, A.V. Muratov, Yu.A. Aleshchenko, M.A. Anisimov, N.Yu. Shitsevalova, S.E. Polovets, V.B. Filipov, V.V. Voronov, and N.E. Sluchanko, Collective infrared excitation in rare-earth $Gd_xLa_{1-x}B_6$ hexaborides, Phys. Rev. B **100**, 104302 (2019).
19. N. B. Bolotina, A. P. Dudka, O. N. Khrykina, V.S. Mironov, "Crystal structure of dodecaborides: complexity in simplicity", in *Rare-Earth Borides*, edited by D. S. Inosov (Jenny Stanford Publishing, Singapore, 2021), Chap.3, pp. 293-330.
20. A. N. Azarevich, N. B. Bolotina, O. N. Khrykina, A. V. Bogach, K. M. Krasikov, A. Yu. Tsvetkov, S. Yu. Gavrilkin, V. V. Voronov, S. Gabani, K. Flachbart, A. V. Kuznetsov, N. E. Sluchanko, Two-gap superconductor $ZrB_{12}$ with dynamic stripes and charge density waves: Crystal structure, physical properties and pairing mechanism, https://doi.org/10.48550/arXiv.2505.23424
21. I. A. Troyan, D. V. Semenok, A. G. Kvashnin, A. V. Sadakov, O. A. Sobolevskiy, V. M. Pudalov, A. G. Ivanova, V. B. Prakapenka, E. Greenberg, A. G. Gavriliuk, I. S. Lyubutin, V. V. Struzhkin, A. Bergara, I. Errea, R. Bianco, M. Calandra, F. Mauri, L. Monacelli, R. Akashi, and A. R. Oganov, "Anomalous High-Temperature Superconductivity in $YH_6$", Adv. Mater. **33**, 2006832, (2021).
22. I. A. Troyan, D. V. Semenok, A. G. Ivanova, A. G. Kvashnin, D. Zhou, A. V. Sadakov, O. A. Sobolevskiy, V. M. Pudalov, I. S. Lyubutin, A. R. Oganov, "High-temperature superconductivity in hydrides", Phys. Usp. **192**, 799–813, (2022).
23. D. Sun, V. S. Minkov, S. Mozaffari, Y. Sun, Y. Ma, S. Chariton, V. B. Prakapenka, M. I. Eremets, L. Balicas, F. F. Balakirev, "High-temperature superconductivity on the verge of a structural instability in lanthanum superhydride", Nat. Comm. **12**, 6863, (2021).
24. N.B. Bolotina, O.N. Khrykina, A.N. Azarevich, N.Yu. Shitsevalova, V.B. Filipov, S.Yu. Gavrilkin, A.Yu. Tsvetkov, S. Gabáni, K. Flachbart, V.V. Voronov, N.E. Sluchanko, "Low temperature singularities of electron density in a two-gap superconductor $ZrB_{12}$", Solid St. Sci. **142**, 107245, (2023).
25. S. Thakur et al., D. Biswas, N. Sahadev, P. K. Biswas, G. Balakrishnan, K. Maiti, "Complex spectral evolution in a BCS superconductor $ZrB_{12}$", Sci. Rep. **3**, 3342, (2013).
26. N.B. Bolotina, A.P. Dudka, O.N. Khrykina, V.N. Krasnorussky, N.Yu. Shitsevalova, V.B. Filipov, and N.E. Sluchanko, The lower symmetry electron-density distribution and the charge transport anisotropy in cubic dodecaboride $LuB_{12}$, J. Phys. Condens. Matter **30**, 265402 (2018).
27. B. P. Gorshunov, E. S. Zhukova, G. A. Komandin, V. I. Torgashev, A. V. Muratov, Yu. A. Aleshchenko, S. V. Demishev, N. Yu. Shitsevalova, V. B. Filipov, N. E. Sluchanko, "Collective Infrared Excitation in $LuB_{12}$ Cage-Glass", JETP Letters **107**, 100 (2018).
28. N. Shitsevalova, Chapter 1. Crystal chemistry and crystal growth of rare-earth borides in rare-earth borides, D. S. Inosov, Ed., Singapore: Jenny Stanford Publishing Pte. Ltd., 2021, pp. 1–244.
29. S.-I. Kimura, T. Namba, S. Kunii, T. Kusuya, Interband Optical Spectra of Rare-Earth Hexaborides, J. Phys. Soc. Jpn 59, 3388-3392 (1990).

30. H. Okamura, S. Kimura, H. Shinozaki, T. Namba, F. Iga, N. Shimizu, T. Takabatake, Phys. Rev. B **58**, R7496 (1998).
31. M. Dressel and G. Grüner, *Electrodynamics of Solids* (Cambridge University Press, Cambridge, UK, 2002).
32. N.E. Sluchanko, A.N. Azarevich, A.V. Bogach, N.B. Bolotina, V.V. Glushkov, S.V. Demishev, A.P. Dudka, O.N. Khrykina, V.B. Filipov, N.Y. Shitsevalova, G.A. Komandin, A.V. Muratov, Y.A. Aleshchenko, E.S. Zhukova, and B.P. Gorshunov, Observation of dynamic charge stripes in $Tm_{0.19}Yb_{0.81}B_{12}$ at the metal−insulator transition, J. Phys.: Condens. Matter **31**, 065604 (2019).
33. E. S. Zhukova, A.Melentyev, B. P. Gorshunov, A. V. Muratov, Yu. A. Aleshchenko, A. N. Azarevich, K. M. Krasikov, N. Shitsevalova, V. Filipov, N. E. Sluchanko, Low-temperature infrared spectroscopy of $Tm_{0.19}Yb_{0.81}B_{12}$ dodecaboride with metal-insulator transition and dynamic charge stripes, J. Phys.: Condens. Matter **34**, 465603 (2022).
34. A. Azarevich, N. Bolotina, O. Khrykina, A. Bogach, E. Zhukova, B. Gorshunov, A. Melentev, Z. Bedran, A. Alyabyeva, M. Belyanchikov, V. Voronov, N.Yu. Shitsevalova, V.B. Filipov, N. Sluchanko, Evidence of electronic phase separation in the strongly correlated semiconductor $YbB_{12}$, Chinese Phys. Lett. **39**, 127302 (2022).
35. N.E. Sluchanko, E.S. Zhukova, L.N. Alyabyeva, B.P. Gorshunov, A.V. Muratov, Yu.A. Aleshchenko, A.N. Azarevich, M.A. Anisimov, N.Yu. Shitsevalova, S.E. Polovets, and V.B. Filipov, Collective and quasi-local modes in the optical spectra of $YB_6$ and $YbB_6$ hexaborides with Jahn–Teller structural instability, J. Exp. Theor. Phys. **136**, 148 (2023).
36. A. V. Sokolov, *Optical properties of metals* (American Elsevier, New York, 1967).
37. V. A. Gasparov, I. Sheikin, F. Levy, J. Teyssier, G. Santi, "Study of the Fermi Surface of $ZrB_{12}$ Using the de Haas-van Alphen Effect", Phys. Rev. Lett. **101**, 097006, (2008).
38. J. Teyssier, A. B. Kuzmenko, D. van der Marel, F. Mersiglio, Optical study of electronic structure and electron-phonon coupling in $ZrB_{12}$, Phys. Rev. B **75**, 134503 (2007).
39. S. Kimura, T. Namba, S. Kunii, T. Kasuya, Low-energy optical excitations in rare-earth hexaborides, Phys. Rev. B **50**, 1406 (1994).
40. A. Kakizaki, A. Harasawa, T. Ishii, T. Kashiwakura, A. Kamata, S. Kunii, Electronic Structure of $CeB_6$ Studied by 3d XPS and High-Resolution 4d-4f Resonant Photoemission, J. Phys. Soc. Jpn. **64**, 302 (1995)